\documentclass[nofootinbib,prd]{revtex4}%
\usepackage{amsmath}
\usepackage{amsfonts}
\usepackage{amssymb}
\usepackage{graphicx}%
\begin{document}
\title{Self sustained traversable wormholes?}
\author{Remo Garattini}
\email{Remo.Garattini@unibg.it}
\affiliation{Universit\`{a} degli Studi di Bergamo, Facolt\`{a} di Ingegneria, Viale
Marconi 5, 24044 Dalmine (Bergamo) ITALY.}

\begin{abstract}
We compute the graviton one loop contribution to a classical energy in a
\textit{traversable} wormhole background. Such a contribution is evaluated by
means of a variational approach with Gaussian trial wave functionals. A zeta
function regularization is involved to handle with divergences. A
renormalization procedure is introduced and the finite one loop energy is
considered as a \textit{self-consistent} source for the traversable wormhole.

\end{abstract}
\maketitle

\section{Introduction}

A wormhole can be represented by two asymptotically flat regions joined by a
bridge. To exist, they must satisfy the Einstein field equations, like a black
hole. While black holes are generally well accepted astrophysical objects,
wormhole are yet to be discovered. One very simple and at the same time
fundamental example of wormhole is represented by the Schwarzschild solution
of the Einstein's field equations. One of the prerogatives of a wormhole is
its ability to connect two distant points in space-time. In this amazing
perspective, it is immediate to recognize the possibility of traveling
crossing wormholes as a short-cut in space and time. Unfortunately, although
there is no direct experience, a Schwarzschild wormhole does not possess this
property. It is for this reason that in a pioneering work Morris and
Thorne\cite{MT} and subsequently Morris, Thorne and Yurtsever\cite{MTY}
studied a class of wormholes termed \textquotedblleft\textit{traversable}%
\textquotedblright. Unfortunately, the traversability is accompanied by
unavoidable violations of null energy conditions, namely, the matter threading
the wormhole's throat has to be \textquotedblleft\textit{exotic}%
\textquotedblright. Classical matter satisfies the usual energy conditions.
Therefore, it is likely that wormholes must belong to the realm of
semiclassical or perhaps a possible quantum theory of the gravitational field.
Since a complete theory of quantum gravity has yet to come, it is important to
approach this problem semiclassically. On this ground, Hochberg, Popov and
Sushkov considered a self-consistent solution of the semiclassical Einstein
equations corresponding to a Lorentzian wormhole coupled to a quantum scalar
field\cite{HPS}. On the other hand, Khusnutdinov and Sushkov fixed their
attention to the computation of the ground state of a massive scalar field in
a wormhole background. They tried to see if a self-consistent solution
restricted to the energy component appears in this configuration\cite{KS}.
Motivated by these works, we are interested to repeat the approach of
Ref.\cite{KS} but using gravitons instead of scalars. In particular, we wish
to study the one loop contribution of the gravitons to the total energy. This
is quite similar to compute the Casimir energy on a fixed background. It is
known that, for different physical systems, Casimir energy is negative. This
is exactly the feature that the exotic matter should possess. In particular,
we conjecture that quantum fluctuations can support the traversability as
effective source of the semiclassical Einstein's equations. The rest of the
paper is structured as follows, in section \ref{p0} we define the effective
Einstein equations, in section \ref{p1} we introduce the traversable wormhole
metric, in section \ref{p2} we give some of the basic rules to perform the
functional integration and we define the Hamiltonian approximated up to second
order, in section \ref{p3} we study the spectrum of the spin-two operator
acting on transverse traceless tensors, in section \ref{p4} we regularize and
renormalize the one loop energy contribution and we speculate about
self-consistency of the result. We summarize and conclude in section \ref{p5}.

\section{The effective Einstein equations}

\label{p0}We begin with a look at the classical Einstein equations%
\begin{equation}
G_{\mu\nu}=\kappa T_{\mu\nu}, \label{Gmunuc}%
\end{equation}
where $T_{\mu\nu}$ is the stress-energy tensor, $G_{\mu\nu}$ is the Einstein
tensor and $\kappa=8\pi G$. Consider a separation of the metric into a
background part, $\bar{g}_{\mu\nu}$, and a perturbation, $h_{\mu\nu}$,%
\begin{equation}
g_{\mu\nu}=\bar{g}_{\mu\nu}+h_{\mu\nu}.
\end{equation}
The Einstein tensor $G_{\mu\nu}$ can also be divided into a part describing
the curvature due to the background geometry and that due to the perturbation,%
\begin{equation}
G_{\mu\nu}\left(  g_{\alpha\beta}\right)  =G_{\mu\nu}\left(  \bar{g}%
_{\alpha\beta}\right)  +\Delta G_{\mu\nu}\left(  \bar{g}_{\alpha\beta
},h_{\alpha\beta}\right)  ,
\end{equation}
where, in principle $\Delta G_{\mu\nu}\left(  \bar{g}_{\alpha\beta}%
,h_{\alpha\beta}\right)  $ is a perturbation series in terms of $h_{\mu\nu}$.
In the context of semiclassical gravity, Eq.$\left(  \ref{Gmunuc}\right)  $
becomes%
\begin{equation}
G_{\mu\nu}=\kappa\left\langle T_{\mu\nu}\right\rangle ^{ren}, \label{Gmunus}%
\end{equation}
where $\left\langle T_{\mu\nu}\right\rangle ^{ren}$ is the renormalized
expectation value of the stress-energy tensor operator of the quantized field.
If the matter field source is absent, nothing prevents us from defining an
effective stress-energy tensor for the fluctuations as\footnote{Note that our
approach is very close to the gravitational \textit{geon} considered by
Anderson and Brill\cite{AndersonBrill}. The relevant difference is in the
averaging procedure.}%
\begin{equation}
\left\langle T_{\mu\nu}\right\rangle ^{ren}=-\frac{1}{\kappa}\left\langle
\Delta G_{\mu\nu}\left(  \bar{g}_{\alpha\beta},h_{\alpha\beta}\right)
\right\rangle ^{ren}.
\end{equation}
From this point of view, the equation governing quantum fluctuations behaves
as a backreaction equation. If we fix our attention to the energy component of
the Einstein field equations, we need to introduce a time-like unit vector
$u^{\mu}$ such that $u\cdot u=-1$. Then the semi-classical Einstein's
equations $\left(  \ref{Gmunus}\right)  $ projected on the constant time
hypersurface $\Sigma$ become%
\begin{equation}
G_{\mu\nu}\left(  \bar{g}_{\alpha\beta}\right)  u^{\mu}u^{\nu}=\kappa
\left\langle T_{\mu\nu}u^{\mu}u^{\nu}\right\rangle ^{ren}=-\left\langle \Delta
G_{\mu\nu}\left(  \bar{g}_{\alpha\beta},h_{\alpha\beta}\right)  u^{\mu}u^{\nu
}\right\rangle ^{ren}.
\end{equation}
To further proceed, it is convenient to consider the associated tensor density
and integrate over $\Sigma$. This leads to\footnote{Details on sign
conventions and decomposition of the Einstein tensor can be found in Apeendix
\ref{app1}}%
\begin{equation}
\frac{1}{2\kappa}\int_{\Sigma}d^{3}x\sqrt{^{3}\bar{g}}G_{\mu\nu}\left(
\bar{g}_{\alpha\beta}\right)  u^{\mu}u^{\nu}=-\int_{\Sigma}d^{3}%
x\mathcal{H}^{\left(  0\right)  }=-\frac{1}{2\kappa}\int_{\Sigma}d^{3}%
x\sqrt{^{3}\bar{g}}\left\langle \Delta G_{\mu\nu}\left(  \bar{g}_{\alpha\beta
},h_{\alpha\beta}\right)  u^{\mu}u^{\nu}\right\rangle ^{ren}, \label{inteq}%
\end{equation}
where%
\begin{equation}
\mathcal{H}^{\left(  0\right)  }=\frac{2\kappa}{\sqrt{^{3}\bar{g}}}G_{ijkl}%
\pi^{ij}\pi^{kl}-\frac{1}{2\kappa}\sqrt{^{3}\bar{g}}R^{\left(  3\right)  }
\label{hdens}%
\end{equation}
is the background field super-hamiltonian and $G_{ijkl}$ is the DeWitt super
metric. Thus the fluctuations in the Einstein tensor are, in this context, the
fluctuations of the hamiltonian. To compute the expectation value of the
perturbed Einstein tensor in the transverse-traceless sector, we use a
variational procedure with gaussian wave functionals. In practice, the right
hand side of Eq.$\left(  \ref{inteq}\right)  $ will be obtained by expanding%
\begin{equation}
E_{wormhole}=\frac{\left\langle \Psi\left\vert H_{\Sigma}\right\vert
\Psi\right\rangle }{\left\langle \Psi|\Psi\right\rangle }=\frac{\left\langle
\Psi\left\vert H_{\Sigma}^{\left(  0\right)  }+H_{\Sigma}^{\left(  1\right)
}+H_{\Sigma}^{\left(  2\right)  }+\ldots\right\vert \Psi\right\rangle
}{\left\langle \Psi|\Psi\right\rangle }%
\end{equation}
and retaining only quantum fluctuations contributing to the effective stress
energy tensor. $H_{\Sigma}^{\left(  i\right)  }$ represents the hamiltonian
approximated to the $i^{th}$ order in $h_{ij}$ and $\Psi$ is a \textit{trial
wave functional} of the gaussian form. Then Eq.$\left(  \ref{inteq}\right)  $
becomes%
\begin{equation}
H_{\Sigma}^{\left(  0\right)  }=\int_{\Sigma}d^{3}x\mathcal{H}^{\left(
0\right)  }=-\frac{\left\langle \Psi\left\vert H_{\Sigma}^{\left(  1\right)
}+H_{\Sigma}^{\left(  2\right)  }+\ldots\right\vert \Psi\right\rangle
}{\left\langle \Psi|\Psi\right\rangle }. \label{flham}%
\end{equation}
The chosen background to compute the quantity contained in Eq.$\left(
\ref{inteq}\right)  $ will be that of a traversable wormhole.

\section{The traversable wormhole metric}

\label{p1}In Schwarzschild coordinates, the traversable wormhole metric can be
cast into the form%
\begin{equation}
ds^{2}=-\exp\left(  -2\phi\left(  r\right)  \right)  dt^{2}+\frac{dr^{2}%
}{1-\frac{b\left(  r\right)  }{r}}+r^{2}\left[  d\theta^{2}+\sin^{2}\theta
d\varphi^{2}\right]  . \label{metric}%
\end{equation}
where $\phi\left(  r\right)  $ is called the redshift function, while
$b\left(  r\right)  $ is called the shape function. Proper radial distance is
related to the shape function by%
\begin{equation}
l\left(  r\right)  =\pm\int_{r_{t}}^{r}\frac{dr^{\prime}}{\sqrt{1-\frac
{b_{\pm}\left(  r^{\prime}\right)  }{r^{\prime}}}}, \label{prd}%
\end{equation}
where the plus (minus) sign is related to the upper (lower) part of the
wormhole or universe. Two coordinate patches are required, each one covering
the range $\left[  r_{t},+\infty\right)  $. Each patch covers one universe,
and the two patches join at $r_{t}$, the throat of the wormhole defined by%
\begin{equation}
r_{t}=\min\left\{  r\left(  l\right)  \right\}  . \label{throat}%
\end{equation}
Instead of Eq.$\left(  \ref{metric}\right)  $, we consider the special case
where $\phi\equiv0$ and $b\left(  r\right)  =r_{t}^{2}/r$. Then, we get%
\begin{equation}
ds^{2}=-dt^{2}+\frac{dr^{2}}{1-\frac{r_{t}^{2}}{r^{2}}}+r^{2}\left[
d\theta^{2}+\sin^{2}\theta d\varphi^{2}\right]  \label{me1}%
\end{equation}
and with the help of Eq.$\left(  \ref{prd}\right)  $, we obtain%
\begin{equation}
l\left(  r\right)  =\pm\int_{r_{t}}^{r}\frac{dr^{\prime}}{\sqrt{1-\frac
{r_{t}^{2}}{r^{\prime2}}}}=\pm\sqrt{r^{2}-r_{t}^{2}}\qquad\Longrightarrow
\qquad r^{2}=l^{2}+r_{t}^{2}%
\end{equation}
and Eq.$\left(  \ref{me1}\right)  $ becomes%
\begin{equation}
ds^{2}=-dt^{2}+dl^{2}+\left(  r_{t}^{2}+l^{2}\right)  \left[  d\theta^{2}%
+\sin^{2}\theta d\varphi^{2}\right]  . \label{me2}%
\end{equation}
The new coordinate $l$ covers the range $-\infty<l<+\infty$. The constant time
hypersurface $\Sigma$ is an Einstein-Rosen bridge with wormhole topology
$S^{2}\times R^{1}$. The Einstein-Rosen bridge defines a bifurcation surface
dividing $\Sigma$ in two parts denoted by $\Sigma_{+}$ and $\Sigma_{-}$. To
concretely compute Eq.$\left(  \ref{flham}\right)  $, we consider on the slice
$\Sigma$ deviations from the wormhole metric of the type
\begin{equation}
g_{ij}=\bar{g}_{ij}+h_{ij}, \label{i3}%
\end{equation}
where $g_{ij}$ is extracted from the line element $\left(  \ref{me2}\right)  $
whose form becomes%
\begin{equation}
ds^{2}=-dt^{2}+g_{ij}dx^{i}dx^{j}.
\end{equation}

\section{Energy Density Calculation in Schr\"{o}dinger Representation}

\label{p2}In order to compute the quantity%
\begin{equation}
-\int_{\Sigma}d^{3}x\sqrt{^{3}\bar{g}}\left\langle \Delta G_{\mu\nu}\left(
\bar{g}_{\alpha\beta},h_{\alpha\beta}\right)  u^{\mu}u^{\nu}\right\rangle
^{ren}, \label{deltagmn}%
\end{equation}
we consider the right hand side of Eq.$\left(  \ref{flham}\right)  $. Since
$H_{\Sigma}^{\left(  1\right)  }$ is linear in $h_{ij}$ and $h$, the
corresponding gaussian integral disappears and since%
\begin{equation}
\frac{\sqrt{^{3}g}}{\kappa}G_{\mu\nu}\left(  g_{\alpha\beta}\right)  u^{\mu
}u^{\nu}=-\mathcal{H}, \label{GmunuH}%
\end{equation}
it is clear that the hamiltonian expansion in Eq.$\left(  \ref{flham}\right)
$ does not coincide with the averaged expanded Einstein tensor of Eq.$\left(
\ref{deltagmn}\right)  $ because Eq.$\left(  \ref{GmunuH}\right)  $ involves a
tensor density. Therefore, the correct setting is%
\begin{equation}
\int_{\Sigma}d^{3}x\sqrt{^{3}\bar{g}}\left\langle \Delta G_{\mu\nu}\left(
\bar{g}_{\alpha\beta},h_{\alpha\beta}\right)  u^{\mu}u^{\nu}\right\rangle
^{ren}=\int_{\Sigma}d^{3}x\sqrt{^{3}\bar{g}}\frac{\left\langle \Psi\left\vert
\mathcal{H}^{\left(  2\right)  }-\sqrt{^{3}g}^{\left(  2\right)  }%
\mathcal{H}^{\left(  0\right)  }\right\vert \Psi\right\rangle }{\left\langle
\Psi|\Psi\right\rangle }, \label{GmunuH2}%
\end{equation}
where $\sqrt{^{3}g}^{\left(  2\right)  }$ is the second order expanded tensor
density weight. Following the same procedure of Refs.\cite{Remo,Remo1}, the
potential part of the right hand side of Eq.$\left(  \ref{GmunuH2}\right)  $
becomes%
\begin{equation}
\int_{\Sigma}d^{3}x\sqrt{\bar{g}}\left[  -\frac{1}{4}h\triangle h+\frac{1}%
{4}h^{li}\triangle h_{li}-\frac{1}{2}h^{ij}\nabla_{l}\nabla_{i}h_{j}^{l}%
+\frac{1}{2}h\nabla_{l}\nabla_{i}h^{li}-\frac{1}{2}h^{ij}R_{ia}h_{j}^{a}%
+\frac{1}{2}hR_{ij}h^{ij}+\frac{1}{4}Rh^{ij}h_{ij}-\frac{1}{4}hRh\right]  .
\end{equation}
The term%
\begin{equation}
\int_{\Sigma}d^{3}x\sqrt{\bar{g}}\left[  \frac{1}{4}Rh^{ij}h_{ij}-\frac{1}%
{4}hRh\right]  ,
\end{equation}
makes the difference between the hamiltonian expansion and the Einstein tensor
expansion. To explicitly make calculations, we need an orthogonal
decomposition for both $\pi_{ij\text{ }}$and $h_{ij}$ to disentangle gauge
modes from physical deformations. We define the inner product%

\begin{equation}
\left\langle h,k\right\rangle :=\int_{\Sigma}\sqrt{g}G^{ijkl}h_{ij}\left(
x\right)  k_{kl}\left(  x\right)  d^{3}x,
\end{equation}
by means of the inverse WDW metric $G_{ijkl}$, to have a metric on the space
of deformations, i.e. a quadratic form on the tangent space at h, with%

\begin{equation}%
\begin{array}
[c]{c}%
G^{ijkl}=(g^{ik}g^{jl}+g^{il}g^{jk}-2g^{ij}g^{kl})\text{.}%
\end{array}
\end{equation}
The inverse metric is defined on co-tangent space and it assumes the form%

\begin{equation}
\left\langle p,q\right\rangle :=\int_{\Sigma}\sqrt{g}G_{ijkl}p^{ij}\left(
x\right)  q^{kl}\left(  x\right)  d^{3}x\text{,}%
\end{equation}
so that%

\begin{equation}
G^{ijnm}G_{nmkl}=\frac{1}{2}\left(  \delta_{k}^{i}\delta_{l}^{j}+\delta
_{l}^{i}\delta_{k}^{j}\right)  .
\end{equation}
Note that in this scheme the ``inverse metric'' is actually the WDW metric
defined on phase space. The desired decomposition on the tangent space of
3-metric deformations\cite{BergerEbin,York} is:%

\begin{equation}
h_{ij}=\frac{1}{3}hg_{ij}+\left(  L\xi\right)  _{ij}+h_{ij}^{\bot}
\label{p21a}%
\end{equation}
where the operator $L$ maps $\xi_{i}$ into symmetric tracefree tensors%

\begin{equation}
\left(  L\xi\right)  _{ij}=\nabla_{i}\xi_{j}+\nabla_{j}\xi_{i}-\frac{2}%
{3}g_{ij}\left(  \nabla\cdot\xi\right)  .
\end{equation}
Thus the inner product between three-geometries becomes
\[
\left\langle h,h\right\rangle :=\int_{\Sigma}\sqrt{g}G^{ijkl}h_{ij}\left(
x\right)  h_{kl}\left(  x\right)  d^{3}x=
\]%
\begin{equation}
\int_{\Sigma}\sqrt{g}\left[  -\frac{2}{3}h^{2}+\left(  L\xi\right)
^{ij}\left(  L\xi\right)  _{ij}+h^{ij\bot}h_{ij}^{\bot}\right]  . \label{p21b}%
\end{equation}
With the orthogonal decomposition in hand we can define the trial wave
functional
\begin{equation}
\Psi\left\{  h_{ij}\left(  \overrightarrow{x}\right)  \right\}  =\mathcal{N}%
\exp\left\{  -\frac{1}{4l_{p}^{2}}\left[  \left\langle hK^{-1}h\right\rangle
_{x,y}^{\bot}+\left\langle \left(  L\xi\right)  K^{-1}\left(  L\xi\right)
\right\rangle _{x,y}^{\Vert}+\left\langle hK^{-1}h\right\rangle _{x,y}%
^{Trace}\right]  \right\}  ,
\end{equation}
where $\mathcal{N}$ is a normalization factor. We are interested in
perturbations of the physical degrees of freedom. Thus we fix our attention
only to the TT tensor sector reducing therefore the previous form into
\begin{equation}
\Psi\left\{  h_{ij}\left(  \overrightarrow{x}\right)  \right\}  =\mathcal{N}%
\exp\left\{  -\frac{1}{4}\left\langle hK^{-1}h\right\rangle _{x,y}^{\bot
}\right\}  .
\end{equation}
Therefore to calculate the energy density, we need to know the action of some
basic operators on $\Psi\left[  h_{ij}\right]  $. The action of the operator
$h_{ij}$ on $|\Psi\rangle=\Psi\left[  h_{ij}\right]  $ is realized by
\begin{equation}
h_{ij}\left(  x\right)  |\Psi\rangle=h_{ij}\left(  \overrightarrow{x}\right)
\Psi\left\{  h_{ij}\right\}  .
\end{equation}
The action of the operator $\pi_{ij}$ on $|\Psi\rangle$, in general, is%

\begin{equation}
\pi_{ij}\left(  x\right)  |\Psi\rangle=-i\frac{\delta}{\delta h_{ij}\left(
\overrightarrow{x}\right)  }\Psi\left\{  h_{ij}\right\}  .
\end{equation}
The inner product is defined by the functional integration:
\begin{equation}
\left\langle \Psi_{1}\mid\Psi_{2}\right\rangle =\int\left[  \mathcal{D}%
h_{ij}\right]  \Psi_{1}^{\ast}\left\{  h_{ij}\right\}  \Psi_{2}\left\{
h_{kl}\right\}  ,
\end{equation}
and by applying previous functional integration rules, we obtain the
expression of the one-loop-like Hamiltonian form for TT (traceless and
transverse) deformations
\begin{equation}
H_{\Sigma}^{\bot}=\frac{1}{4}\int_{\Sigma}d^{3}x\sqrt{g}G^{ijkl}\left[
\left(  16\pi G\right)  K^{-1\bot}\left(  x,x\right)  _{ijkl}+\frac{1}{\left(
16\pi G\right)  }\left(  \triangle_{2}\right)  _{j}^{a}K^{\bot}\left(
x,x\right)  _{iakl}\right]  . \label{p22}%
\end{equation}
The propagator $K^{\bot}\left(  x,x\right)  _{iakl}$ comes from a functional
integration and it can be represented as
\begin{equation}
K^{\bot}\left(  \overrightarrow{x},\overrightarrow{y}\right)  _{iakl}:=%
%TCIMACRO{\dsum _{\tau}}%
%BeginExpansion
{\displaystyle\sum_{\tau}}
%EndExpansion
\frac{h_{ia}^{\left(  \tau\right)  \bot}\left(  \overrightarrow{x}\right)
h_{kl}^{\left(  \tau\right)  \bot}\left(  \overrightarrow{y}\right)
}{2\lambda\left(  \tau\right)  },
\end{equation}
where $h_{ia}^{\left(  \tau\right)  \bot}\left(  \overrightarrow{x}\right)  $
are the eigenfunctions of $\triangle_{2}$, whose eigenvalues will be denoted
with $\omega^{2}\left(  \tau\right)  $. $\tau$ denotes a complete set of
indices and $\lambda\left(  \tau\right)  $ are a set of variational parameters
to be determined by the minimization of Eq.$\left(  \ref{p22}\right)  $. The
expectation value of $H^{\bot}$ is easily obtained by inserting the form of
the propagator into Eq.$\left(  \ref{p22}\right)  $%
\begin{equation}
E\left(  \lambda_{i}\right)  =\frac{1}{4}%
%TCIMACRO{\dsum _{\tau}}%
%BeginExpansion
{\displaystyle\sum_{\tau}}
%EndExpansion%
%TCIMACRO{\dsum _{i=1}^{2}}%
%BeginExpansion
{\displaystyle\sum_{i=1}^{2}}
%EndExpansion
\left[  \left(  16\pi G\right)  \lambda_{i}\left(  \tau\right)  +\frac
{\omega_{i}^{2}\left(  \tau\right)  }{\left(  16\pi G\right)  \lambda
_{i}\left(  \tau\right)  }\right]  .
\end{equation}
By minimizing with respect to the variational function $\lambda_{i}\left(
\tau\right)  $, we obtain the total one loop energy for TT tensors%
\begin{equation}
E^{TT}=\frac{1}{4}%
%TCIMACRO{\dsum _{\tau}}%
%BeginExpansion
{\displaystyle\sum_{\tau}}
%EndExpansion
\left[  \sqrt{\omega_{1}^{2}\left(  \tau\right)  }+\sqrt{\omega_{2}^{2}\left(
\tau\right)  }\right]  . \label{e1loop}%
\end{equation}
The above expression makes sense only for $\omega_{i}^{2}\left(  \tau\right)
>0$, $i=1,2$. The meaning of $\omega_{i}^{2}$ will be clarified in the next
section. Coming back to Eq.$\left(  \ref{flham}\right)  $, we observe that the
value of the wormhole energy on the chosen background is%
\begin{equation}
\int_{\Sigma}d^{3}x\mathcal{H}^{\left(  0\right)  }=-\frac{1}{16\pi G}%
\int_{\Sigma}d^{3}x\sqrt{\bar{g}}R^{\left(  3\right)  }=\frac{\pi r_{t}}{2G}.
\label{ecl}%
\end{equation}
and to one loop the self-consistent equation for TT tensors becomes%
\begin{equation}
\frac{\pi r_{t}}{2G}=-E^{TT}. \label{sceq}%
\end{equation}
Note that the self-consistency on the hamiltonian as a reversed sign with
respect to the energy component of the Einstein field equations. This means
that an eventual stable point for the hamiltonian is an unstable point for the
effective energy momentum tensor and vice versa.

\section{The transverse traceless (TT) spin 2 operator for the traversable
wormhole and the W.K.B. approximation}

\label{p3}In this section, we evaluate the one loop energy expressed by
Eq.$\left(  \ref{e1loop}\right)  $. To this purpose, we begin with the
operator describing gravitons propagating on the background $\left(
\ref{me2}\right)  $. The Lichnerowicz operator in this particular metric is
defined by%
\begin{equation}
\left(  \triangle_{2}h^{TT}\right)  _{i}^{j}:=-\left(  \triangle_{T}%
h^{TT}\right)  _{i}^{j}+2\left(  Rh^{TT}\right)  _{i}^{j}-R\left(
h^{TT}\right)  _{i}^{j}, \label{spin2}%
\end{equation}
where the transverse-traceless (TT) tensor for the quantum fluctuation is
obtained by the following decomposition%
\begin{equation}
h_{i}^{j}=h_{i}^{j}-\frac{1}{3}\delta_{i}^{j}h+\frac{1}{3}\delta_{i}%
^{j}h=\left(  h^{T}\right)  _{i}^{j}+\frac{1}{3}\delta_{i}^{j}h.
\end{equation}
This implies that $\left(  h^{T}\right)  _{i}^{i}=0$. The transversality
condition is applied on $\left(  h^{T}\right)  _{i}^{j}$ and becomes
$\nabla_{j}\left(  h^{T}\right)  _{i}^{j}=0$. Thus%
\begin{equation}
-\left(  \triangle_{T}h^{TT}\right)  _{i}^{j}=-\triangle_{S}\left(
h^{TT}\right)  _{i}^{j}+\frac{6}{r^{2}}\left(  1-\frac{b\left(  r\right)  }%
{r}\right)  , \label{tlap}%
\end{equation}
where $\triangle_{S}$ is the scalar curved Laplacian, whose form is%
\begin{equation}
\triangle_{S}=\left(  1-\frac{b\left(  r\right)  }{r}\right)  \frac{d^{2}%
}{dr^{2}}+\left(  \frac{4r-b^{\prime}\left(  r\right)  r-3b\left(  r\right)
}{2r^{2}}\right)  \frac{d}{dr}-\frac{L^{2}}{r^{2}} \label{slap}%
\end{equation}
and $R_{j\text{ }}^{a}$ is the mixed Ricci tensor whose components are:
\begin{equation}
R_{i}^{a}=\left\{  \frac{b^{\prime}\left(  r\right)  }{r^{2}}-\frac{b\left(
r\right)  }{r^{3}},\frac{b^{\prime}\left(  r\right)  }{2r^{2}}+\frac{b\left(
r\right)  }{2r^{3}},\frac{b^{\prime}\left(  r\right)  }{2r^{2}}+\frac{b\left(
r\right)  }{2r^{3}}\right\}  .
\end{equation}
The scalar curvature is%
\begin{equation}
R=R_{i}^{j}\delta_{j}^{i}=2\frac{b^{\prime}\left(  r\right)  }{r^{2}}%
\end{equation}
We are therefore led to study the following eigenvalue equation
\begin{equation}
\left(  \triangle_{2}h^{TT}\right)  _{i}^{j}=\omega^{2}h_{i}^{j} \label{p31}%
\end{equation}
where $\omega^{2}$ is the eigenvalue of the corresponding equation. In doing
so, we follow Regge and Wheeler in analyzing the equation as modes of definite
frequency, angular momentum and parity\cite{RW}. In particular, our choice for
the three-dimensional gravitational perturbation is represented by its
even-parity form%

\begin{equation}
h_{ij}^{even}\left(  r,\vartheta,\phi\right)  =diag\left[  H\left(  r\right)
\left(  1-\frac{b\left(  r\right)  }{r}\right)  ^{-1},r^{2}K\left(  r\right)
,r^{2}\sin^{2}\vartheta L\left(  r\right)  \right]  Y_{lm}\left(
\vartheta,\phi\right)  , \label{pert}%
\end{equation}
with%
\begin{equation}
\left\{
\begin{array}
[c]{c}%
H\left(  r\right)  =h_{1}^{1}\left(  r\right)  -\frac{1}{3}h\left(  r\right)
\\
K\left(  r\right)  =h_{2}^{2}\left(  r\right)  -\frac{1}{3}h\left(  r\right)
\\
L\left(  r\right)  =h_{3}^{3}\left(  r\right)  -\frac{1}{3}h\left(  r\right)
\end{array}
\right.  .
\end{equation}
From the transversality condition we obtain $h_{2}^{2}\left(  r\right)
=h_{3}^{3}\left(  r\right)  $. Then $K\left(  r\right)  =L\left(  r\right)  $.
For a generic value of the angular momentum $L$, representation $\left(
\ref{pert}\right)  $ joined to Eq.$\left(  \ref{tlap}\right)  $ lead to the
following system of PDE's%

\begin{equation}
\left\{
\begin{array}
[c]{c}%
\left(  -\triangle_{l}+2\left(  \frac{b^{\prime}\left(  r\right)  }{r^{2}%
}-\frac{b\left(  r\right)  }{r^{3}}-\frac{b^{\prime}\left(  r\right)  }{r^{2}%
}\right)  \right)  H\left(  r\right)  =\omega_{1,l}^{2}H\left(  r\right) \\
\\
\left(  -\triangle_{l}+2\left(  \frac{b^{\prime}\left(  r\right)  }{2r^{2}%
}+\frac{b\left(  r\right)  }{2r^{3}}-\frac{b^{\prime}\left(  r\right)  }%
{r^{2}}\right)  \right)  K\left(  r\right)  =\omega_{2,l}^{2}K\left(
r\right)
\end{array}
\right.  , \label{p33}%
\end{equation}
where $\triangle_{l}$ is%

\begin{equation}
\triangle_{l}=\left(  1-\frac{b\left(  r\right)  }{r}\right)  \frac{d^{2}%
}{dr^{2}}+\left(  \frac{4r-b^{\prime}\left(  r\right)  r-3b\left(  r\right)
}{2r^{2}}\right)  \frac{d}{dr}-\frac{l\left(  l+1\right)  }{r^{2}}-\frac
{6}{r^{2}}\left(  1-\frac{b\left(  r\right)  }{r}\right)  . \label{p33a}%
\end{equation}
Defining reduced fields%

\begin{equation}
H\left(  r\right)  =\frac{f_{1}\left(  r\right)  }{r};\qquad K\left(
r\right)  =\frac{f_{2}\left(  r\right)  }{r},
\end{equation}
and passing to the proper geodesic distance from the \textit{throat} of the
bridge defined by%
\begin{equation}
dx=\pm\frac{dr}{\sqrt{1-\frac{b\left(  r\right)  }{r}}},
\end{equation}
the system $\left(  \ref{p33}\right)  $ becomes%

\begin{equation}
\left\{
\begin{array}
[c]{c}%
\left[  -\frac{d^{2}}{dx^{2}}+V_{1}\left(  x\right)  \right]  f_{1}\left(
x\right)  =\omega_{1,l}^{2}f_{1}\left(  x\right) \\
\\
\left[  -\frac{d^{2}}{dx^{2}}+V_{2}\left(  x\right)  \right]  f_{2}\left(
x\right)  =\omega_{2,l}^{2}f_{2}\left(  x\right)
\end{array}
\right.  \label{p34}%
\end{equation}
with
\begin{equation}
\left\{
\begin{array}
[c]{c}%
V_{1}\left(  x\right)  =\frac{l\left(  l+1\right)  }{x^{2}+r_{t}^{2}}%
+U_{1}\left(  x\right) \\
\\
V_{2}\left(  x\right)  =\frac{l\left(  l+1\right)  }{x^{2}+r_{t}^{2}}%
+U_{2}\left(  x\right)
\end{array}
\right.  ,
\end{equation}
and%
\begin{equation}
\left\{
\begin{array}
[c]{c}%
U_{1}\left(  x\right)  =\left[  \frac{6}{\left(  x^{2}+r_{t}^{2}\right)  ^{2}%
}x^{2}-\frac{r_{t}^{2}}{\left(  x^{2}+r_{t}^{2}\right)  ^{2}}\right] \\
\\
U_{2}\left(  x\right)  =\left[  \frac{6}{\left(  x^{2}+r_{t}^{2}\right)  ^{2}%
}x^{2}+\frac{3r_{t}^{2}}{\left(  x^{2}+r_{t}^{2}\right)  ^{2}}\right]
\end{array}
\right.  . \label{potentials}%
\end{equation}
In order to use the WKB approximation, we define two r-dependent radial wave
numbers $k_{1}\left(  x,l,\omega_{1,nl}\right)  $ and $k_{2}\left(
x,l,\omega_{2,nl}\right)  $%
\begin{equation}
\left\{
\begin{array}
[c]{c}%
k_{1}^{2}\left(  x,l,\omega_{1,nl}\right)  =\omega_{1,nl}^{2}-\frac{l\left(
l+1\right)  }{\left(  x^{2}+r_{t}^{2}\right)  }-U_{1}\left(  x\right) \\
\\
k_{2}^{2}\left(  x,l,\omega_{2,nl}\right)  =\omega_{2,nl}^{2}-\frac{l\left(
l+1\right)  }{\left(  x^{2}+r_{t}^{2}\right)  }-U_{2}\left(  x\right)
\end{array}
\right.  . \label{rwn}%
\end{equation}
The number of modes with frequency less than $\omega_{i}$, $i=1,2$, is given
approximately by%
\begin{equation}
\tilde{g}\left(  \omega_{i}\right)  =\int\nu_{i}\left(  l,\omega_{i}\right)
\left(  2l+1\right)  dl,
\end{equation}
where $\nu_{i}\left(  l,\omega_{i}\right)  $, $i=1,2$ is the number of nodes
in the mode with $\left(  l,\omega_{i}\right)  $, such that%
\begin{equation}
\nu_{i}\left(  l,\omega_{i}\right)  =\frac{1}{\pi}\int_{-\infty}^{+\infty
}dx\sqrt{k_{i}^{2}\left(  x,l,\omega_{i}\right)  }.
\end{equation}
Here it is understood that the integration with respect to $x$ and $l$ is
taken over those values which satisfy $k_{i}^{2}\left(  x,l,\omega_{i}\right)
\geq0,$ $i=1,2$. Thus the total one loop energy for TT tensors is given by%
\[
E^{TT}=\frac{1}{4}\sum_{i=1}^{2}\int_{0}^{+\infty}\omega_{i}\frac{d\tilde
{g}\left(  \omega_{i}\right)  }{d\omega_{i}}d\omega_{i}=\sum_{i=1}^{2}%
\int_{-\infty}^{+\infty}dx\left(  x^{2}+r_{t}^{2}\right)  \left[  \frac
{1}{4\pi^{2}}\int_{\sqrt{U_{i}\left(  x\right)  }}^{+\infty}\omega_{i}%
^{2}\sqrt{\omega_{i}^{2}-U_{i}\left(  x\right)  }d\omega_{i}\right]
\]%
\begin{equation}
=\int_{-\infty}^{+\infty}dx\left(  x^{2}+r_{t}^{2}\right)  \left[  \rho
_{1}+\rho_{2}\right]  , \label{tote1loop}%
\end{equation}
where%
\begin{equation}
\left\{
\begin{array}
[c]{c}%
\rho_{1}=\frac{1}{4\pi^{2}}\int_{\sqrt{U_{1}\left(  x\right)  }}^{+\infty
}\omega_{1}^{2}\sqrt{\omega_{1}^{2}-U_{1}\left(  x\right)  }d\omega_{1}\\
\\
\rho_{2}=\frac{1}{4\pi^{2}}\int_{\sqrt{U_{2}\left(  x\right)  }}^{+\infty
}\omega_{2}^{2}\sqrt{\omega_{2}^{2}-U_{2}\left(  x\right)  }d\omega_{2}%
\end{array}
\right.  . \label{edens}%
\end{equation}

\section{One loop energy Regularization and Renormalization}

\label{p4}In this section, we proceed to evaluate the one loop energy. The
method is equivalent to the scattering phase shift method and to the same
method used to compute the entropy in the brick wall model. We use the zeta
function regularization method to compute the energy densities $\rho_{1}$ and
$\rho_{2}$. Note that this procedure is completely equivalent to the
subtraction procedure of the Casimir energy computation where zero point
energy (ZPE) in different backgrounds with the same asymptotic properties is
involved. To this purpose, we introduce the additional mass parameter $\mu$ in
order to restore the correct dimension for the regularized quantities. Such an
arbitrary mass scale emerges unavoidably in any regularization schemes. Then
we have%
\begin{equation}
\rho_{i}\left(  \varepsilon\right)  =\frac{1}{4\pi^{2}}\mu^{2\varepsilon}%
\int_{\sqrt{U_{i}\left(  x\right)  }}^{+\infty}d\omega_{i}\frac{\omega_{i}%
^{2}}{\left(  \omega_{i}^{2}-U_{i}\left(  x\right)  \right)  ^{\varepsilon
-\frac{1}{2}}} \label{zeta}%
\end{equation}
If one of the functions $U_{i}\left(  x\right)  $ is negative, then the
integration has to be meant in the range where $\omega_{i}^{2}+U_{i}\left(
x\right)  \geq0$. In both cases the result of the integration
is\footnote{Details of the calculation can be found in the Appendix
\ref{app2}.}%
\begin{equation}
=-\frac{U_{i}^{2}\left(  x\right)  }{64\pi^{2}}\left[  \frac{1}{\varepsilon
}+\ln\left(  \frac{\mu^{2}}{U_{i}\left(  x\right)  }\right)  +2\ln2-\frac
{1}{2}\right]  , \label{zeta1}%
\end{equation}
where the absolute value has been inserted to take account of the possible
change of sign. Then the total regularized one loop energy is%
\begin{equation}
E^{TT}\left(  r_{t},\varepsilon;\mu\right)  =\int_{-\infty}^{+\infty}dx\left(
x^{2}+r_{t}^{2}\right)  \left[  \left(  \rho_{1}\left(  \varepsilon\right)
+\rho_{2}\left(  \varepsilon\right)  \right)  \right]  .
\end{equation}
The result of the integration over the $x$ coordinate leads to the following
expression%
\begin{equation}
E^{TT}\left(  r_{t},\varepsilon;\mu\right)  =\frac{1}{16\pi}\left[  -\frac
{a}{\varepsilon r_{t}}-\frac{b}{r_{t}}-\frac{c}{r_{t}}\ln\left(  r_{t}%
\mu\right)  \right]  , \label{tote1loopreg}%
\end{equation}
where $a=23.35740218$, $b=276.6026775$ and $c=212.0575042$. Then the self
consistent equation $\left(  \ref{sceq}\right)  $ can be written in the form%
\begin{equation}
\frac{\pi r_{t}}{2G}=\frac{1}{16\pi}\left[  \frac{a}{\varepsilon r_{t}}%
+\frac{b}{r_{t}}+\frac{c}{r_{t}}\ln\left(  r_{t}\mu\right)  \right]  .
\label{sceqtot}%
\end{equation}
In order to deal with finite quantities, we renormalize the divergent energy
by absorbing the singularity in the classical quantity$.$ In particular, we
re-define the bare classical constant $G$%
\begin{equation}
\frac{1}{G}\rightarrow\frac{1}{G_{0}}+\frac{a}{\varepsilon8\pi^{2}r_{t}^{2}}.
\end{equation}
Therefore, the remaining finite value for the effective equation $\left(
\ref{sceqtot}\right)  $ reads%
\begin{equation}
\frac{\pi r_{t}}{2G_{0}}=\frac{1}{16\pi}\left[  \frac{b}{r_{t}}+\frac{c}%
{r_{t}}\ln\left(  r_{t}\mu\right)  \right]  . \label{sceqtoteff}%
\end{equation}
This quantity depends on the arbitrary mass scale $\mu.$ It is appropriate to
use the renormalization group equation to eliminate such a dependence. To this
aim, we impose that\cite{Cherednikov}%
\begin{equation}
\mu\frac{d}{d\mu}\left(  \frac{\pi r_{t}}{2G_{0}\left(  \mu\right)  }\right)
=\mu\frac{d}{d\mu}\left\{  \frac{1}{16\pi}\left[  \frac{b}{r_{t}}+\frac
{c}{r_{t}}\ln\left(  r_{t}\mu\right)  \right]  \right\}  ,
\end{equation}
namely%
\begin{equation}
\frac{\pi r_{t}}{2}\mu\frac{\partial G_{0}^{-1}\left(  \mu\right)  }%
{\partial\mu}-\frac{c}{16\pi r_{t}}=0.
\end{equation}
Solving it we find that the renormalized constant $G_{0}$ should be treated as
a running one in the sense that it varies provided that the scale $\mu$ is
changing
\begin{equation}
\frac{1}{G_{0}\left(  \mu\right)  }=\frac{1}{G_{0}\left(  \mu_{0}\right)
}+K\ln\left(  \frac{\mu}{\mu_{0}}\right)  \qquad\text{or}\qquad G_{0}\left(
\mu\right)  =\frac{G_{0}\left(  \mu_{0}\right)  }{1+G_{0}\left(  \mu
_{0}\right)  K\ln\left(  \frac{\mu}{\mu_{0}}\right)  },\qquad K=\frac{c}%
{8\pi^{2}r_{t}^{2}}; \label{Gmu}%
\end{equation}
where $\mu_{0}$ is the normalization point. We substitute Eq.$\left(
\ref{Gmu}\right)  $ into Eq.$\left(  \ref{sceqtoteff}\right)  $ to find%
\begin{equation}
\frac{\pi}{2G_{0}\left(  \mu_{0}\right)  }=\frac{1}{16\pi}\left[  \frac
{b}{r_{t}^{2}}+\frac{c}{r_{t}^{2}}\ln\left(  r_{t}\mu_{0}\right)  \right]  ,
\label{Nsceqtoteff}%
\end{equation}
where we have divided by $r_{t}$. In order to have only one
solution\footnote{Note that in the paper of Khusnutdinov and Sushkov\cite{KS},
to find only one solution, the minimum of the ground state of the quantized
scalar field has been set equal to the classical energy. In our case, we have
no external fields on a given background. This means that it is not possible
to find a minimum of the one loop gravitons, in analogy with Ref.\cite{KS}.
Moreover the renormalization procedure in Ref.\cite{KS} is completely
independent by the classical term, while in our case it is not. Indeed, thanks
to the self-consistent equation $\left(  \ref{sceq}\right)  $, we can
renormalize the divergent term.}, we find the extremum of the r.h.s. of
Eq.$\left(  \ref{Nsceqtoteff}\right)  $ and we get%
\begin{equation}
\frac{c-2b}{2c}=\ln\left(  \bar{r}_{t}\mu_{0}\right)  \qquad\Longrightarrow
\qquad\bar{r}_{t}=\frac{1}{\mu_{0}}\exp\left(  \frac{c-2b}{2c}\right)
\label{rtmin}%
\end{equation}
and%
\begin{equation}
\frac{1}{G_{0}\left(  \mu_{0}\right)  }=\frac{c\mu_{0}^{2}}{16\pi^{2}}%
\exp\left(  -\frac{c-2b}{c}\right)  . \label{G0mu}%
\end{equation}
With the help of Eqs.$\left(  \ref{rtmin}\right)  $ and $\left(
\ref{G0mu}\right)  $, Eq.$\left(  \ref{Gmu}\right)  $ becomes%
\[
\frac{1}{G_{0}\left(  \mu\right)  }=\frac{1}{G_{0}\left(  \mu_{0}\right)
}+\frac{c}{8\pi^{2}\bar{r}_{t}^{2}}\ln\left(  \frac{\mu}{\mu_{0}}\right)
=\frac{c\mu_{0}^{2}}{16\pi^{2}}\exp\left(  -\frac{c-2b}{c}\right)  \left[
1+2\ln\left(  \frac{\mu}{\mu_{0}}\right)  \right]
\]%
\begin{equation}
=\frac{1}{G_{0}\left(  \mu_{0}\right)  }\left[  1+2\ln\left(  \frac{\mu}%
{\mu_{0}}\right)  \right]  .
\end{equation}
It is straightforward to see that we have a constraint on $\mu/\mu_{0}$.
Indeed we have to choose%
\begin{equation}
\mu>\mu_{0}\exp\left(  -\frac{1}{2}\right)  =.6065306597\mu_{0},
\end{equation}
otherwise $G_{0}\left(  \mu\right)  $ becomes negative. We have now two possibilities:

\begin{enumerate}
\item we identify $G_{0}\left(  \mu_{0}\right)  $ with the squared Planck
length. As a consequence we obtain%
\begin{equation}
\bar{r}_{t}=\frac{1}{4\pi}\sqrt{cG_{0}\left(  \mu_{0}\right)  }%
=1.158822606l_{p}\qquad\Longrightarrow\qquad\mu_{0}=.3860531213m_{p}.
\label{a}%
\end{equation}

\item We identify $\mu_{0}$ with the Planck scale and we get%
\begin{equation}
\bar{r}_{t}=.4473670842l_{p}\qquad\Longrightarrow\qquad\sqrt{G_{0}\left(
\mu_{0}\right)  }=.3860531213l_{p}. \label{b}%
\end{equation}
If we compare our result with that obtained by Khusnutdinov and
Sushkov\cite{KS}, we see that in their work they find that the wormhole radius
is%
\begin{equation}
r_{w}\simeq.0141l_{p} \label{rw}%
\end{equation}
and its mass%
\begin{equation}
m_{w}\simeq11.35m_{p}%
\end{equation}
is trans-planckian. We recall that these values have been obtained by looking
at a massive scalar field quantized in a wormhole background playing the role
of the exotic matter. In our case, we suppose that the graviton quantum
fluctuations play the role of the exotic matter and even if we fix the
renormalization point at the Planck scale as in Ref.\cite{KS}, we find a
radius $\bar{r}_{t}>r_{w}$.
\end{enumerate}

\section{Summary and Conclusions}

\label{p5}Motivated by the works of Anderson and Brill, in searching for
gravitational geons\cite{AndersonBrill} and of Khusnutdinov and
Sushkov\cite{KS}, we have considered the possible existence of a
self-consistent solution of the semiclassical Einstein's equations in a
traversable wormhole background. In particular, we have fixed our attention to
the graviton quantum fluctuations around such a background. The fluctuations,
contained in the perturbed Einstein tensor, play the role of the
\textquotedblleft\textit{exotic}\textquotedblright\ matter considered in
Ref.\cite{KS}. A variational approach with the help of gaussian trial wave
functionals has been used to compute the one loop term. The adopted procedure
is quite close to that of Anderson and Brill except for the averaging process
which, in our case, involves a variational calculation. To handle the
divergencies we have used a zeta function regularization which, in this
context, is formally equivalent to a Casimir energy subtraction procedure. A
renormalization of the Newton's constant $G$ has been performed to absorb the
ultraviolet divergencies. To avoid dependences on the renormalization scale a
renormalization group equation has been computed. While the renormalization
process is not new in the context of semiclassical Einstein field equations,
to our knowledge it is the use of the renormalization group equation that
seems to be unknown, especially concerning self-consistent solutions and
traversability. The interesting point is the obtained consistency with a
wormhole radius $\bar{r}_{t}$ greater than radius $r_{w}$ of Eq.$\left(
\ref{rw}\right)  $. It is likely that this result be correlated with the
graviton themselves. Despite of this, the obtained \textquotedblleft%
\textit{traversability}\textquotedblright\ has to be regarded as in
\textquotedblleft\textit{principle}\textquotedblright\ rather than in
\textquotedblleft\textit{practice}\textquotedblright\ because of the obtained
wormhole radius size. Nevertheless, the calculation is far to be complete
because only gravitons have been taken into account and the traversable
wormhole metric considered is restricted to Eq.$\left(  \ref{me1}\right)  $. A
possible improvement of our evaluation could come from including a
charge\cite{Kim} or by looking at a Schwarzschild-like metric, but even with a
different choice of the wave functionals entering in the variational approach.

\appendix

\section{Einstein equations and the Hamiltonian}

\label{app1}Let us consider the Einstein equations%
\begin{equation}
G_{\mu\nu}=R_{\mu\nu}-\frac{1}{2}g_{\mu\nu}R=\kappa8\pi GT_{\mu\nu}.
\end{equation}
$R_{\mu\nu}$ is the Ricci tensor and $R$ is the scalar curvature. If $u^{\mu}$
is a time-like unit vector such that $g_{\mu\nu}u^{\mu}u^{\nu}=-1,$then the
Einstein tensor $G_{\mu\nu}$ becomes%
\begin{equation}
G_{\mu\nu}u^{\mu}u^{\nu}=R_{\mu\nu}u^{\mu}u^{\nu}-\frac{1}{2}g_{\mu\nu}u^{\mu
}u^{\nu}R=R_{\mu\nu}u^{\mu}u^{\nu}+\frac{1}{2}R. \label{a1}%
\end{equation}
By means of the Gauss-Codazzi equations\cite{HawEll},
\begin{equation}
R=R^{\left(  3\right)  }\pm2R_{\mu\nu}u^{\mu}u^{\nu}\mp K^{2}\pm K_{\mu\nu
}K^{\mu\nu},
\end{equation}
where $K_{\mu\nu}$ is the extrinsic curvature and $R^{\left(  3\right)  }$ is
the three dimensional scalar curvature. For a time-like vector, we take the
lower sign and Eq.$\left(  \ref{a1}\right)  $ becomes%
\begin{equation}
G_{\mu\nu}u^{\mu}u^{\nu}=\frac{1}{2}\left(  R^{\left(  3\right)  }%
+K^{2}-K_{\mu\nu}K^{\mu\nu}\right)  . \label{a2}%
\end{equation}
If the conjugate momentum is defined by%
\begin{equation}
\pi^{\mu\nu}=\frac{\sqrt{^{\left(  3\right)  }g}}{2\kappa}\left(  Kg^{\mu\nu
}-K^{\mu\nu}\right)  ,
\end{equation}
then%
\begin{equation}
K^{2}-K_{\mu\nu}K^{\mu\nu}=\left(  \frac{2\kappa}{\sqrt{^{\left(  3\right)
}g}}\right)  ^{2}\left(  \frac{\pi^{2}}{2}-\pi^{\mu\nu}\pi_{\mu\nu}\right)
\end{equation}
and%
\begin{equation}
\frac{\sqrt{^{\left(  3\right)  }g}}{2\kappa}G_{\mu\nu}u^{\mu}u^{\nu}%
=\frac{\sqrt{^{\left(  3\right)  }g}}{2\kappa}R^{\left(  3\right)  }%
+\frac{2\kappa}{\sqrt{^{\left(  3\right)  }g}}\left(  \frac{\pi^{2}}{2}%
-\pi^{\mu\nu}\pi_{\mu\nu}\right)  =-\mathcal{H}^{\left(  0\right)  },
\end{equation}
namely Eq.$\left(  \ref{hdens}\right)  $ with the reversed sign.

\section{The zeta function regularization}

\label{app2}In this appendix, we report details on computation leading to
expression $\left(  \ref{zeta}\right)  $. We begin with the following integral%
\begin{equation}
\rho\left(  \varepsilon\right)  =\left\{
\begin{array}
[c]{c}%
I_{+}=\mu^{2\varepsilon}\int_{0}^{+\infty}d\omega\frac{\omega^{2}}{\left(
\omega^{2}+U\left(  x\right)  \right)  ^{\varepsilon-\frac{1}{2}}}\\
\\
I_{-}=\mu^{2\varepsilon}\int_{0}^{+\infty}d\omega\frac{\omega^{2}}{\left(
\omega^{2}-U\left(  x\right)  \right)  ^{\varepsilon-\frac{1}{2}}}%
\end{array}
\right.  , \label{rho}%
\end{equation}
with $U\left(  x\right)  >0$.

\subsection{$I_{+}$ computation}

\label{app2a}If we define $t=\omega/\sqrt{U\left(  x\right)  }$, the integral
$I_{+}$ in Eq.$\left(  \ref{rho}\right)  $ becomes%
\[
\rho\left(  \varepsilon\right)  =\mu^{2\varepsilon}U\left(  x\right)
^{2-\varepsilon}\int_{0}^{+\infty}dt\frac{t^{2}}{\left(  t^{2}+1\right)
^{\varepsilon-\frac{1}{2}}}=\frac{1}{2}\mu^{2\varepsilon}U\left(  x\right)
^{2-\varepsilon}B\left(  \frac{3}{2},\varepsilon-2\right)
\]%
\begin{equation}
\frac{1}{2}\mu^{2\varepsilon}U\left(  x\right)  ^{2-\varepsilon}\frac
{\Gamma\left(  \frac{3}{2}\right)  \Gamma\left(  \varepsilon-2\right)
}{\Gamma\left(  \varepsilon-\frac{1}{2}\right)  }=\frac{\sqrt{\pi}}{4}U\left(
x\right)  ^{2}\left(  \frac{\mu^{2}}{U\left(  x\right)  }\right)
^{\varepsilon}\frac{\Gamma\left(  \varepsilon-2\right)  }{\Gamma\left(
\varepsilon-\frac{1}{2}\right)  },
\end{equation}
where we have used the following identities involving the beta function%
\begin{equation}
B\left(  x,y\right)  =2\int_{0}^{+\infty}dt\frac{t^{2x-1}}{\left(
t^{2}+1\right)  ^{x+y}}\qquad\operatorname{Re}x>0,\operatorname{Re}y>0
\end{equation}
related to the gamma function by means of%
\begin{equation}
B\left(  x,y\right)  =\frac{\Gamma\left(  x\right)  \Gamma\left(  y\right)
}{\Gamma\left(  x+y\right)  }.
\end{equation}
Taking into account the following relations for the $\Gamma$-function%
\begin{equation}
\Gamma\left(  \varepsilon-2\right)  =\frac{\Gamma\left(  1+\varepsilon\right)
}{\varepsilon\left(  \varepsilon-1\right)  \left(  \varepsilon-2\right)
},\qquad\Gamma\left(  \varepsilon-\frac{1}{2}\right)  =\frac{\Gamma\left(
\varepsilon+\frac{1}{2}\right)  }{\varepsilon-\frac{1}{2}}, \label{gamma}%
\end{equation}
and the expansion for small $\varepsilon$%
\[
\Gamma\left(  1+\varepsilon\right)  =1-\gamma\varepsilon+O\left(
\varepsilon^{2}\right)  ,\qquad\Gamma\left(  \varepsilon+\frac{1}{2}\right)
=\Gamma\left(  \frac{1}{2}\right)  -\varepsilon\Gamma\left(  \frac{1}%
{2}\right)  \left(  \gamma+2\ln2\right)  +O\left(  \varepsilon^{2}\right)
\]%
\begin{equation}
x^{\varepsilon}=1+\varepsilon\ln x+O\left(  \varepsilon^{2}\right)  ,
\end{equation}
where $\gamma$ is the Euler's constant, we find%
\begin{equation}
\rho\left(  \varepsilon\right)  =-\frac{U^{2}\left(  x\right)  }{16}\left[
\frac{1}{\varepsilon}+\ln\left(  \frac{\mu^{2}}{U\left(  x\right)  }\right)
+2\ln2-\frac{1}{2}\right]  .
\end{equation}

\subsection{$I_{-}$ computation}

\label{app2b}If we define $t=\omega/\sqrt{U\left(  x\right)  }$, the integral
$I_{-}$ in Eq.$\left(  \ref{rho}\right)  $ becomes%
\[
\rho\left(  \varepsilon\right)  =\mu^{2\varepsilon}U\left(  x\right)
^{2-\varepsilon}\int_{0}^{+\infty}dt\frac{t^{2}}{\left(  t^{2}-1\right)
^{\varepsilon-\frac{1}{2}}}=\frac{1}{2}\mu^{2\varepsilon}U\left(  x\right)
^{2-\varepsilon}B\left(  \varepsilon-2,\frac{3}{2}-\varepsilon\right)
\]%
\begin{equation}
\frac{1}{2}\mu^{2\varepsilon}U\left(  x\right)  ^{2-\varepsilon}\frac
{\Gamma\left(  \frac{3}{2}-\varepsilon\right)  \Gamma\left(  \varepsilon
-2\right)  }{\Gamma\left(  -\frac{1}{2}\right)  }=-\frac{1}{4\sqrt{\pi}%
}U\left(  x\right)  ^{2}\left(  \frac{\mu^{2}}{U\left(  x\right)  }\right)
^{\varepsilon}\Gamma\left(  \frac{3}{2}-\varepsilon\right)  \Gamma\left(
\varepsilon-2\right)  ,
\end{equation}
where we have used the following identity involving the beta function%
\begin{equation}
\frac{1}{p}B\left(  1-\nu-\frac{\mu}{p},\nu\right)  =\int_{1}^{+\infty
}dtt^{\mu-1}\left(  t^{p}-1\right)  ^{\nu-1}\qquad p>0,\operatorname{Re}%
\nu>0,\operatorname{Re}\mu<p-p\operatorname{Re}\nu
\end{equation}
and the reflection formula%
\begin{equation}
\Gamma\left(  z\right)  \Gamma\left(  1-z\right)  =-z\Gamma\left(  -z\right)
\Gamma\left(  z\right)
\end{equation}
From the first of Eqs.$\left(  \ref{gamma}\right)  $ and from the expansion
for small $\varepsilon$%
\[
\Gamma\left(  \frac{3}{2}-\varepsilon\right)  =\Gamma\left(  \frac{3}%
{2}\right)  \left(  1-\varepsilon\left(  -\gamma-2\ln2+2\right)  \right)
+O\left(  \varepsilon^{2}\right)
\]%
\begin{equation}
x^{\varepsilon}=1+\varepsilon\ln x+O\left(  \varepsilon^{2}\right)  ,
\end{equation}
we find%
\begin{equation}
\rho\left(  \varepsilon\right)  =-\frac{U^{2}\left(  x\right)  }{16}\left[
\frac{1}{\varepsilon}+\ln\left(  \frac{\mu^{2}}{U\left(  x\right)  }\right)
+2\ln2-\frac{1}{2}\right]  .
\end{equation}

\end{document}